# Development and characterization of single gap glass RPC

Manisha*, V. Bhatnagar, J.S. Shahi, J.B. Singh

*Department of Physics, Panjab University,*
*Chandigarh-160014, India.*

## Abstract

India-based Neutrino Observatory (INO) facility is going to have a 50 kton magnetized Iron CALorimeter (ICAL) detector for precision measurements of neutrino oscillations using atmospheric neutrinos. The proposed ICAL detector will be a stack of magnetized iron plates (acting as target material) interleaved with glass Resistive Plate Chambers (RPCs) as the active detector elements. An RPC is a gaseous detector made up of two parallel electrode plates having high bulk resistivity like that of float glass and bakelite. For the ICAL detector, glass is preferred over bakelite as it does not need any kind of surface treatment to achieve better surface uniformity and also the cost of associated electronics is reduced. Under the detector R&D efforts for the proposed glass RPC detector, a few glass RPCs of 1m X 1m dimension are fabricated procuring glass of ~ 2 mm thickness from one of the Indian glass manufacturers (Asahi). In the present paper, we report the characterization of RPC based on leakage current study, muon detection efficiency and noise rate studies with varying gas compositions.

* Corresponding author : manisha@pu.ac.in

## 1. Introduction

India–based Neutrino Observatory (INO) [1] is the proposed underground facility with an overburden of ~ 1 km to be built at Pottipuram in Bodi West hills in the Western Ghats of India. The Iron CALorimeter (ICAL) detector is one of the biggest atmospheric neutrino detectors which will be established at the INO site. The ICAL will be having three modules with dimensions of ~ 48 m X 16 m X 14.5 m. Each ICAL module will be a stack of 151 layers of iron plates (5.6 cm thick) magnetized with an average magnetic field of ~ 1.5 T. These iron layers will be interleaved with ~ 10,000 glass RPCs (2 m X 2 m) in each module [2]. The primary goal of INO–ICAL is to improve significantly the measurements of neutrino oscillation parameters as compared to the earlier measurements [3, 4]. RPCs [5] are gaseous detectors (pioneered during 1980s) made up of two parallel electrode plates having high bulk resistivity similar to those of glass and bakelite. These detectors have a two pick-up panels based readout system. These pick-up panels are placed orthogonally on each side of RPC gap, to extract the hit information in X and Y coordinates due to passage of charged particle through the RPC. As the ICAL detector will use ~ 30,000 glass RPCs [6], therefore it is necessary to make dedicated efforts for their characterization. In the present paper, leakage current studies with different gas compositions, muon detection efficiency and noise rate studies are reported for a fabricated RPC. Uniformity studies for detection efficiency and noise rate at different locations of the RPC are also reported.

## 2. Fabrication of the glass RPC

Here, we briefly describe the fabrication process of glass RPC, which has been reported in detail by Santonico et al. [6, 7]. Two glass plates having dimensions of 1 m X 1 m X 2 mm are used to make the RPC electrodes. These plates are kept apart by polycarbonate button spacers having 10 mm diameter and 2 mm thickness to maintain uniform separation of 2 mm between the plates. Firstly glass plates are well cleaned with distilled water and subsequently with alcohol. Polycarbonate edge spacers (used to seal the glass RPC gas gap) are glued on one of the glass plates (along with gas nozzles). Once these glued edge spacers are dry enough, button spacers are glued with the help of a mylar template on this glass plate. Then the upper glass plate is glued carefully by gently placing it on the attached button spacers on the lower glass plate. The glass RPC gas gap is first subjected to a gas leak test. Once gas leak rate (*dP/dt* ) is estimated, then it is subjected to a spacer test. Spacer test is performed by gently pressing all the glued button spacers with thumb pressure. As a button spacer is pressed, change in pressure is seen in the form of a peak. If any of the peaks overshoot, it means corresponding button spacer has popped up meaning it is no longer attached to the glass plate. Plots of the leak test and the spacer test are shown in Figure 1. Leak rate is estimated to be ~ $10^{-4}$ millibar/second. If the leak rate is $10^{-3}$ millibar/second [8], then the RPC gas gap is re-glued. Number of peaks in the spacer test plot correspond to the number of button spacers glued between the two glass plates. Both the outer surfaces of gas gap are coated with graphite paint. Surface resistivity measurements of the graphite coated glass RPC gap are performed along both X and Y dimensions. These results are shown as contour plots in Figure 2. All the measured resistivity values have been found to be in the range 0.1 – 0.5 MΩ/□ . The surface resistivity values in the range 0.1 - 1 MΩ/□ have to be achieved for the uniform application of applied voltage [9]. Final assembly of the RPC is done by attaching two pick-up panels on both the sides (each one for X and Y) to the graphite coated surfaces of RPC gap using kapton tape (highly insulating). Pick-up panels are made up of thin copper foils pasted on one side of a plastic honeycomb structure and an aluminium ground plane (shown in Figure 3) on the other side. Each pick-up panel has 32 strips etched out of copper foils having a width 2.8 cm. Distance between two adjacent strips is 2 mm. For insulation purpose, mylar sheets are used between the pick-up panel and the graphite coated RPC gas gap (Figure 3).

## 3. Cosmic ray muon telescope setup

A cosmic ray muon telescope setup is used for the characterization of assembled glass RPC. A telescope assembly consists of 4 scintillator paddles marked as SP1, SP2, SP3 and SP4. The paddles SP1, SP2 are placed below the assembled RPC (parallel to the X readout pick-up panel) and SP3, SP4 are placed above. SP1, SP2 have the dimensions of 20 cm X 84.5 cm and SP3, SP4 which decide the size of the telescope coincidence window have the dimensions of 2.8 cm X 34.5 cm. This way, strip width of the RPC. A discriminated ANDed (4 fold coincidence) signal of these 4 scintillator paddles SP1, SP2, SP3 and SP4 is considered as the master trigger for the RPC testing. Master trigger is latched in coincidence with an RPC strip signal, which forms a 5 fold coincidence trigger. Such a telescope setup is also known as a coincidence setup. Further for the characterization of an RPC, we have used NIM and CAMAC based Data Acquisition (DAQ) system. Counts are recorded using a scalar unit for a time span of 10 minutes at regular intervals for efficiency and noise rate studies. Delay induced due to the signal (analog and digital) carrying cables is taken care by adjusting the cable lengths.

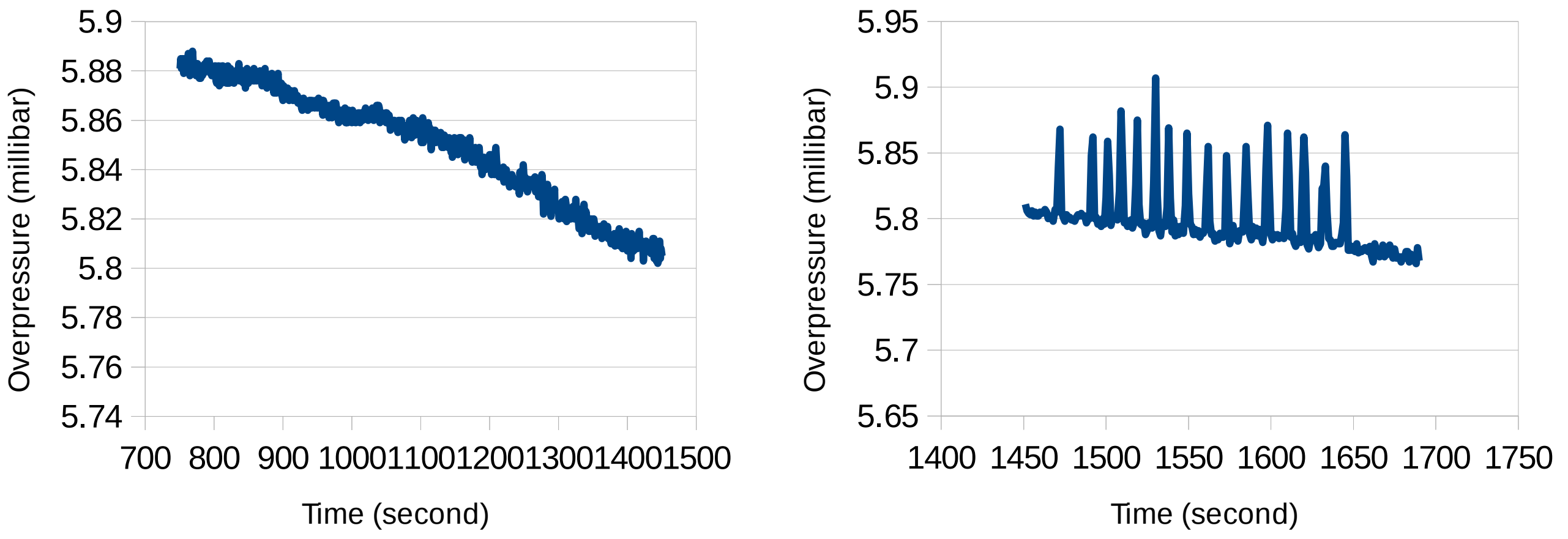

Figure 1: Leak test plot (left), spacer test plot (right) for the glass RPC gap.

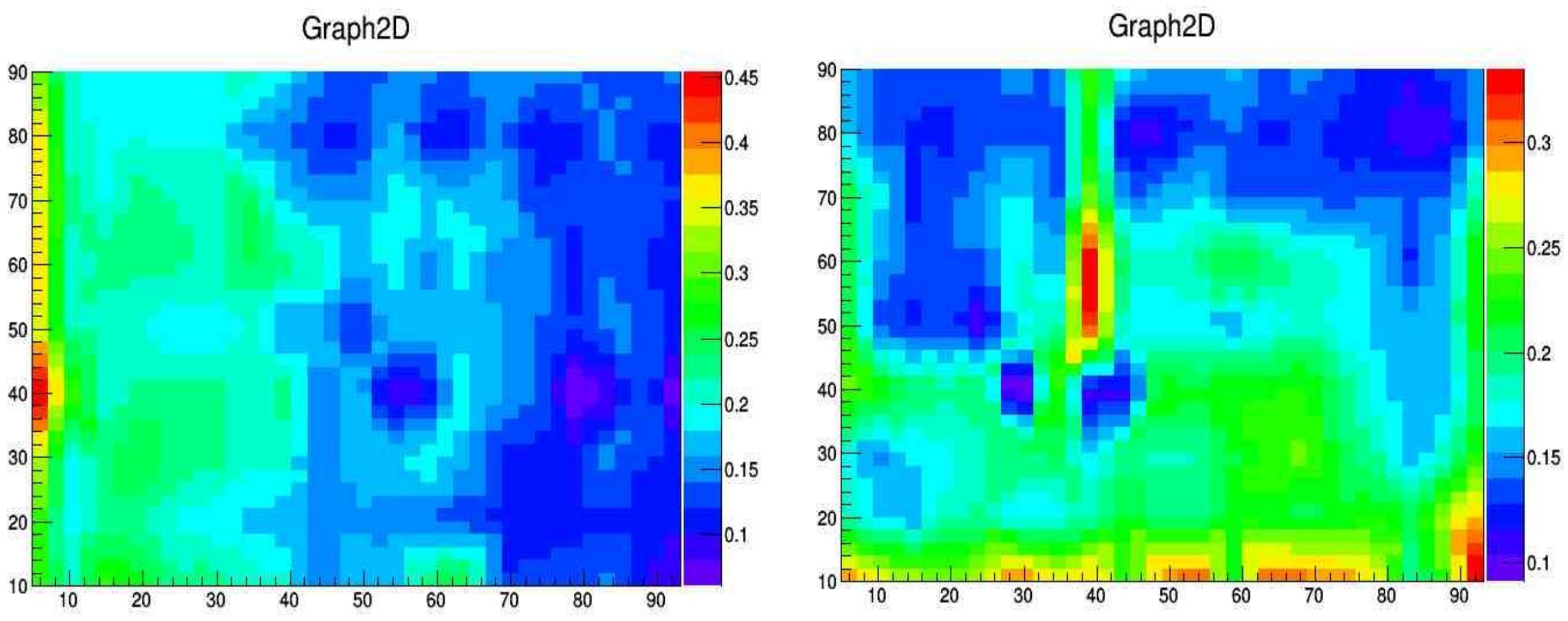

Figure 2: Contour plots of horizontal resistivity measurements (left), vertical resistivity measurements (right). Scale on the right side of each plot shows the resistivity values in MΩ. Glass dimensions along X and Y axes are in cm.

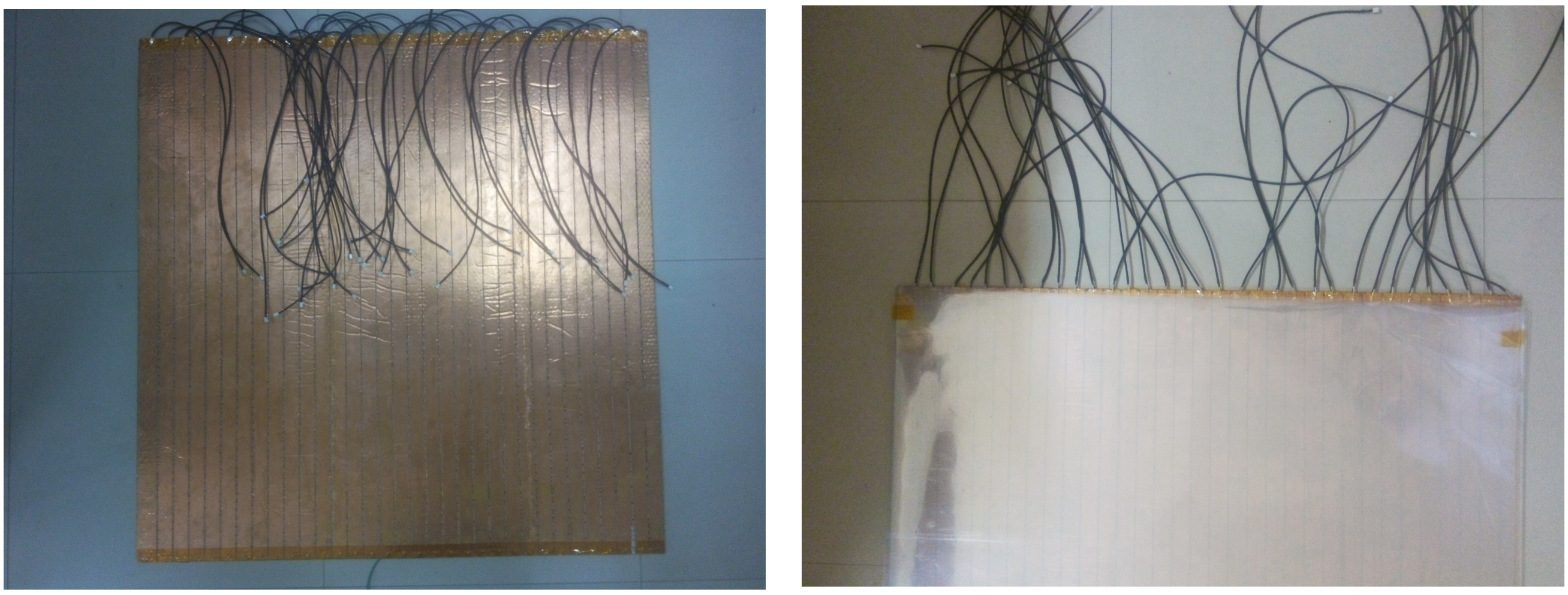

Figure 3: Pick-up panel (left), mylar sheets fixed to pick-up panel (right).

## 4. Detector Characterization

A VI characteristics of an RPC detector comes from the variation of leakage current (I) vs. applied voltage (V) for different gas operating compositions (as mentioned in Table 1), which have not been reported in literature so far. Gases included in these compositions are selected by taking care of following features : high gain, low working voltage, linearity, high rate capability, $N_p$ (number of primary electron pairs per cm), $N_t$ (total number of electron - ion pairs per cm), high absorption coefficient to UV light and a minimum probability of detector ageing etc. [10, 11]. The VI characteristics are the base benchmark for optimizing the operating gas composition. An efficiency and the noise rate studies are performed - to check performance of an RPC and the extent of operational uniformity. More details are given in the following subsections. For all the reported studies, gas flow rate is kept at 1 LPH (liter per hour). Laboratory ambient conditions are monitored round the clock and kept under control with temperature at 20–21℃, the relative humidity within 32-36% . Another important step is to flush the RPC gas gap extensively with nitrogen to remove any residual traces of humidity [12].

Table 1: Different compositions of the gases R134A, $SF_6$, Ar, $C_4H_{10}$ used for the RPC characterization.

| Sr. No. | Freon (R134A) | Isobutane ($C_4H_{10}$) | $SF_6$ | Argon(Ar) |
|---|---|---|---|---|
| 1[St] Composition | 95.2 % | 4.5 % | 0.3 % | - |
| 2[nd] Composition | 95.5 % | 4.5 % | - | - |
| 3[rd] Composition | 100 % | - | - | - |
| 4[th] Composition | - | - | - | 100 % |
| 5[th] Composition | 62 % | 8 % | - | 30 % |

## 4.1 Leakage current

For selecting an optimized gas composition, the variation of leakage current (I) vs. RPC operating voltage (V) is measured for various gas compositions (Table 1). CAEN SY2527 (Universal Multichannel High Voltage Power Supply) is used for providing high voltage to the RPC electrodes. The variation of leakage current vs. voltage is shown in Figure 4. 1[st] gas composition clearly shows the minimum leakage current as SF6 has a quenching effect and the maximum leakage is seen for the 3[rd] gas composition as Freon has a high primary ionization [7]. Furthermore we see from Figure 4 that the 2[nd] gas composition is also good for operating an RPC with low leakage current, therefore to operate the RPC with a small leakage current, the 1[st] and the 2[nd] gas compositions are used for further characterization studies (Figure 5).

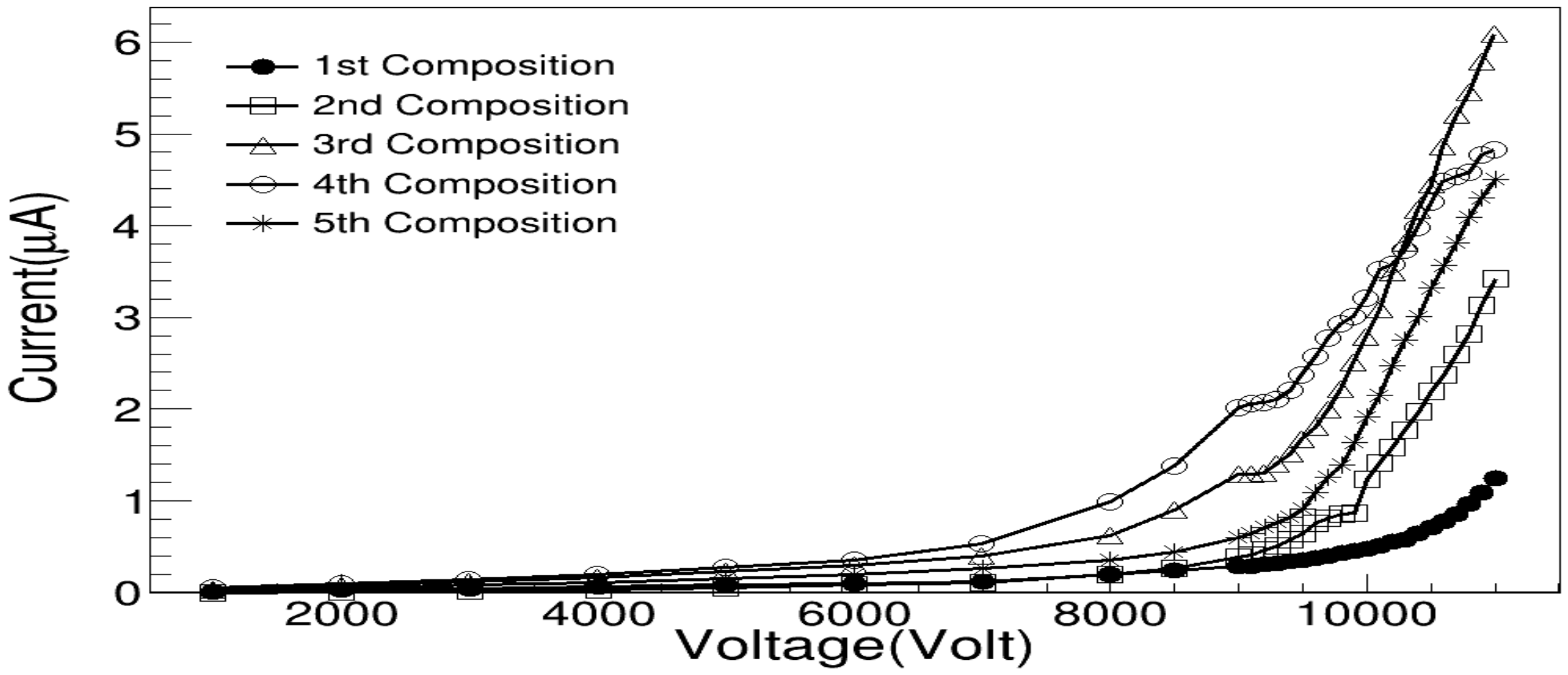

Figure 4: Leakage current vs. voltage characteristics for the different gas compositions.

## 4.2 Efficiency and noise rate

For the 1st and the 2nd gas compositions as mentioned in Table 1, the efficiency and noise rate studies are performed for various readout strips of the RPC. The efficiency and noise rate studies of 4th, 17th and 29th strips are shown in Figure 6 - 8 respectively. The efficiency of all the above mentioned strips of the RPC is ~ 96 - 97% for both the 1st and 2nd gas compositions [13, 14], which is not much affected by the observed noise rate in the range 9 - 18 *Hz/cm²*. For all the tested RPC strips, efficiency increases significantly after 9000 V and reaches the plateau value corresponding to ~ 96 - 97% from 9700 - 9800 V onward. So, we infer that 9700 – 10500 V is the relevant operation region for the RPC. Some unexpected fluctuations are observed in the noise rate of 17th strip around 10000 V, variation in manual graphite coating can be one of the possible reasons for it. Throughout the characterization process, the measured leakage current is found to be in the range 1 – 2 μA (Figure 5).

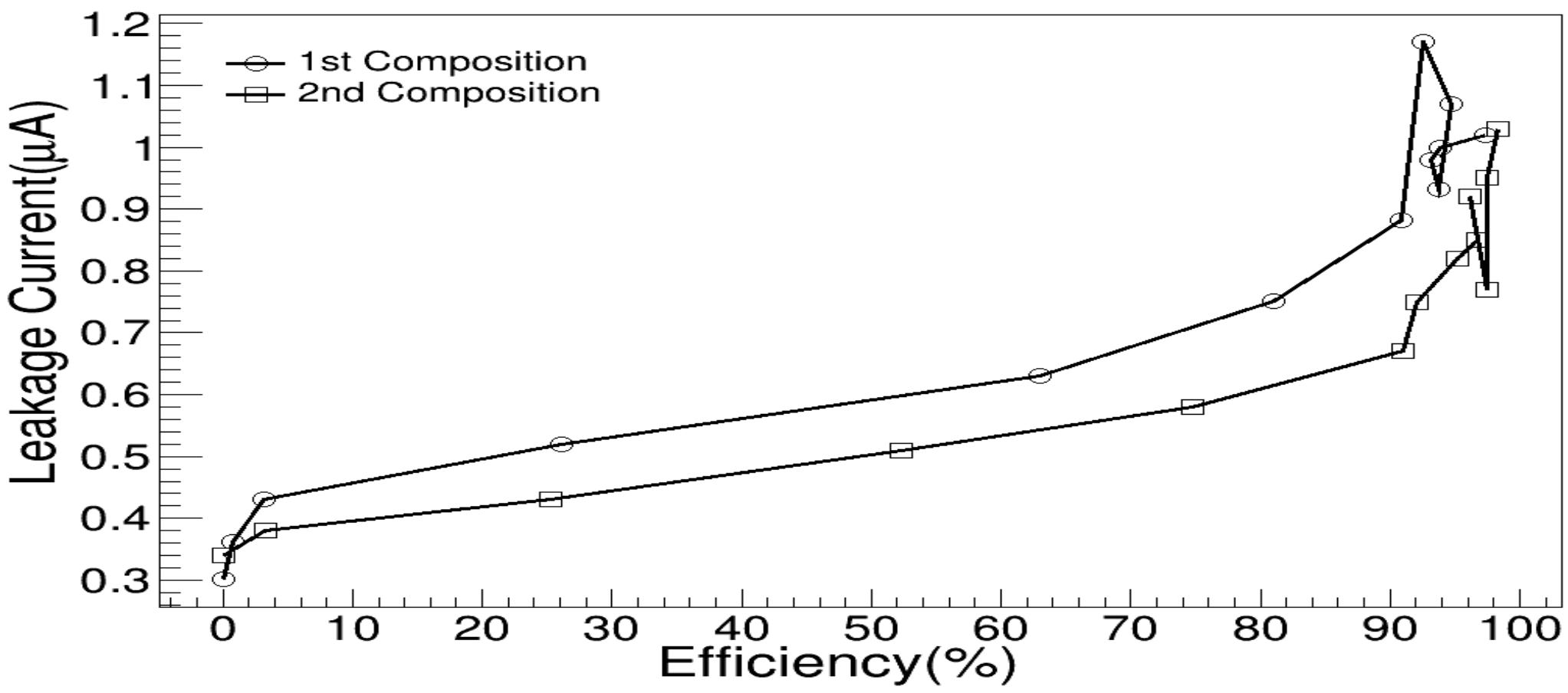

Figure 5 : Leakage current vs. efficiency characteristics.

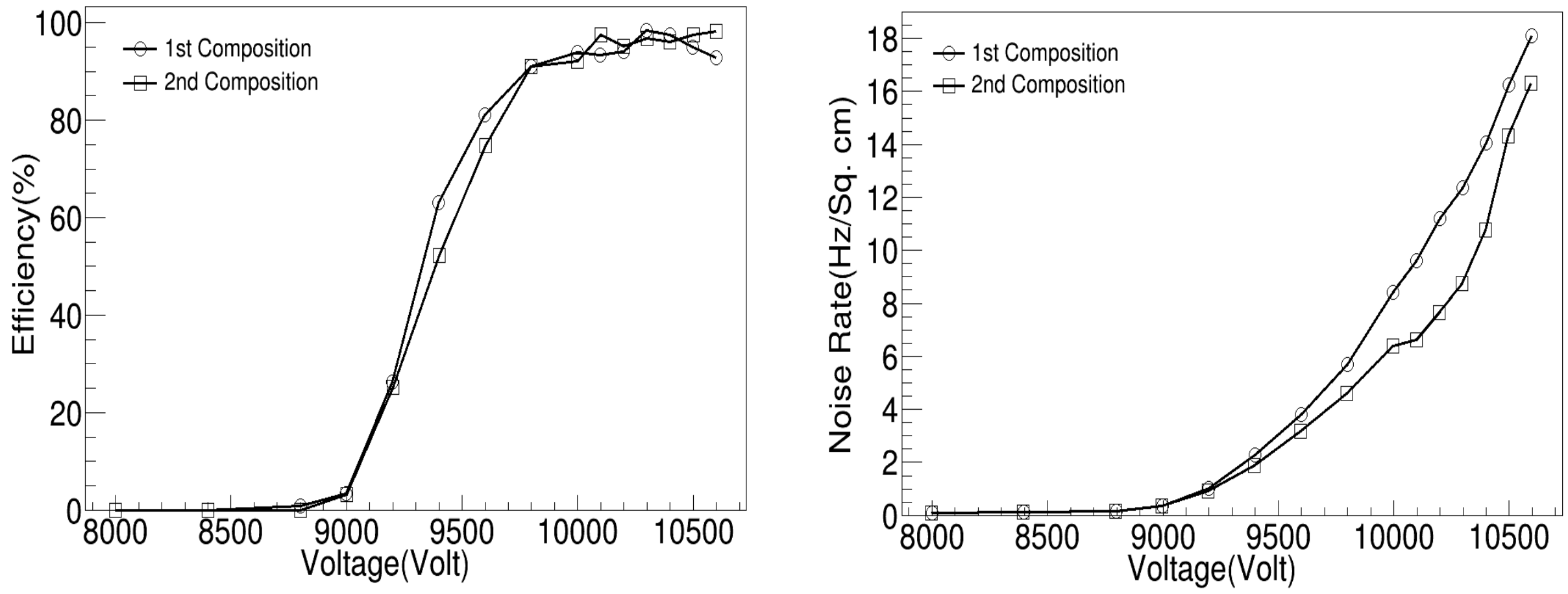

Figure 6: Efficiency vs. voltage characteristics (left), noise rate vs. voltage characteristics (right) for 4th strip.

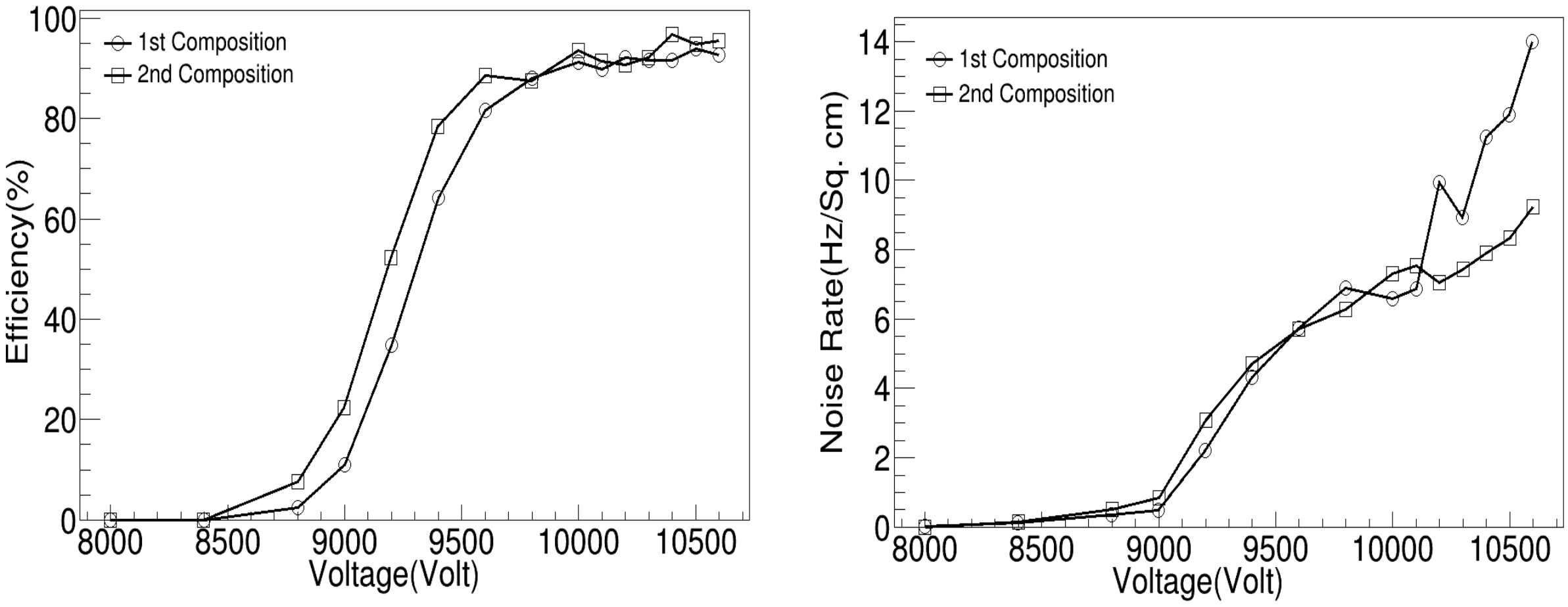

Figure 7: Efficiency vs. voltage characteristics (left), noise rate vs. voltage characteristics (right) for 17th strip.

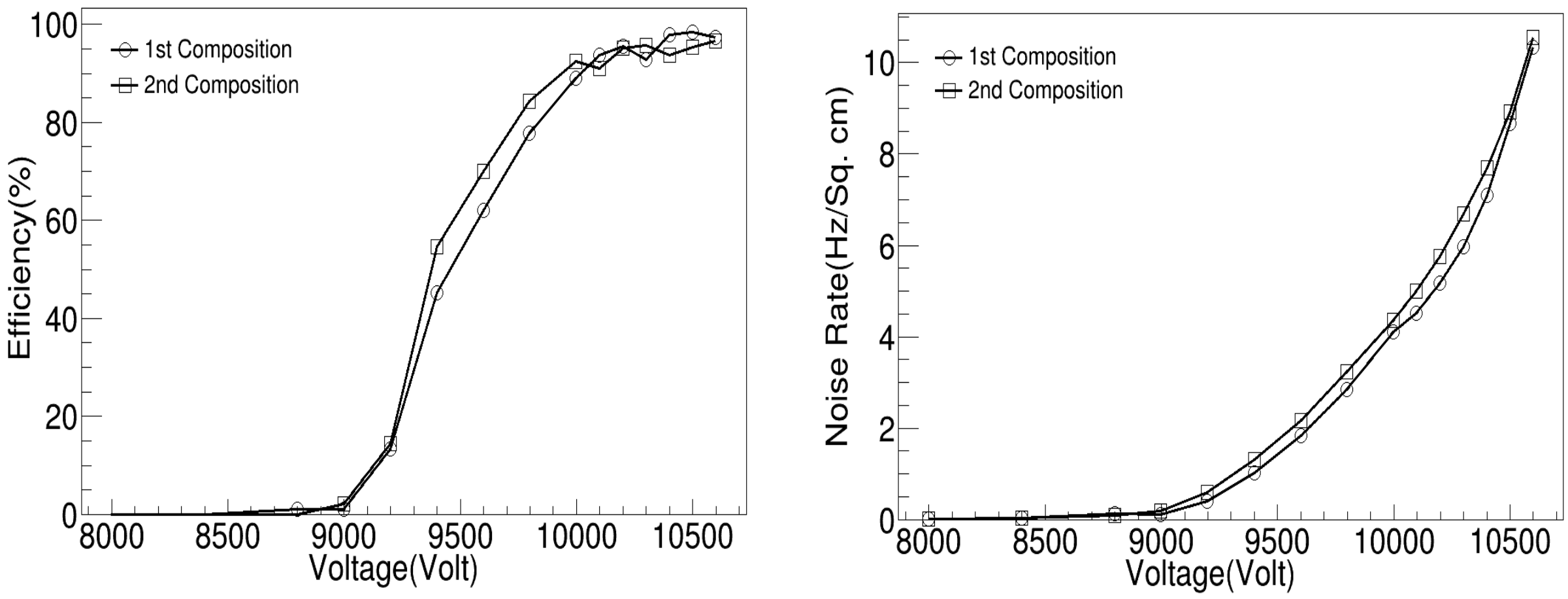

Figure 8: Efficiency vs. voltage characteristics (left), noise rate vs. voltage characteristics (right) for 29th strip.

## 4.3 Uniformity studies

To test uniformity in the efficiency and the noise rate throughout the RPC for the 1st and 2nd gas compositions, corresponding studies are performed at 5 different locations for 5 different strips (4th, 11th, 17th, 23rd and 29th) of the RPC. The RPCs being the active detector element of the ICAL are required to have a uniform and stable performance over the maximum region of coverage.The choice of these strips have been drawn neither a priori nor posteriori. There are total 32 strips out of which strips, 1st and 32nd being the edge strips cover an uncoated portion of the electrode plates on which graphite coating was not done to avoid the external discharge. Therefore, out of remaining 30 strips, efficiency and noise rate measurements are done at 5 different positions on the RPC moving from one end to the other end (spanning the active area).

In addition to this, discriminator threshold values are kept fixed as mentioned earlier (Section 3). The uniformity studies plots are shown in Figures 12, 13 respectively. For the 1st gas composition, the efficiency for these strips is ~ 98% and noise rate is in the range of ~ 10 – 18 *Hz/cm²*. For the 2nd gas composition, the efficiency is in the range of ~ 96 – 98% and the noise rate is in the range of ~ 10 – 16 *Hz/cm²*. Thus, we have observed consistent results for the uniformity in the efficiency and noise rate throughout the RPC.

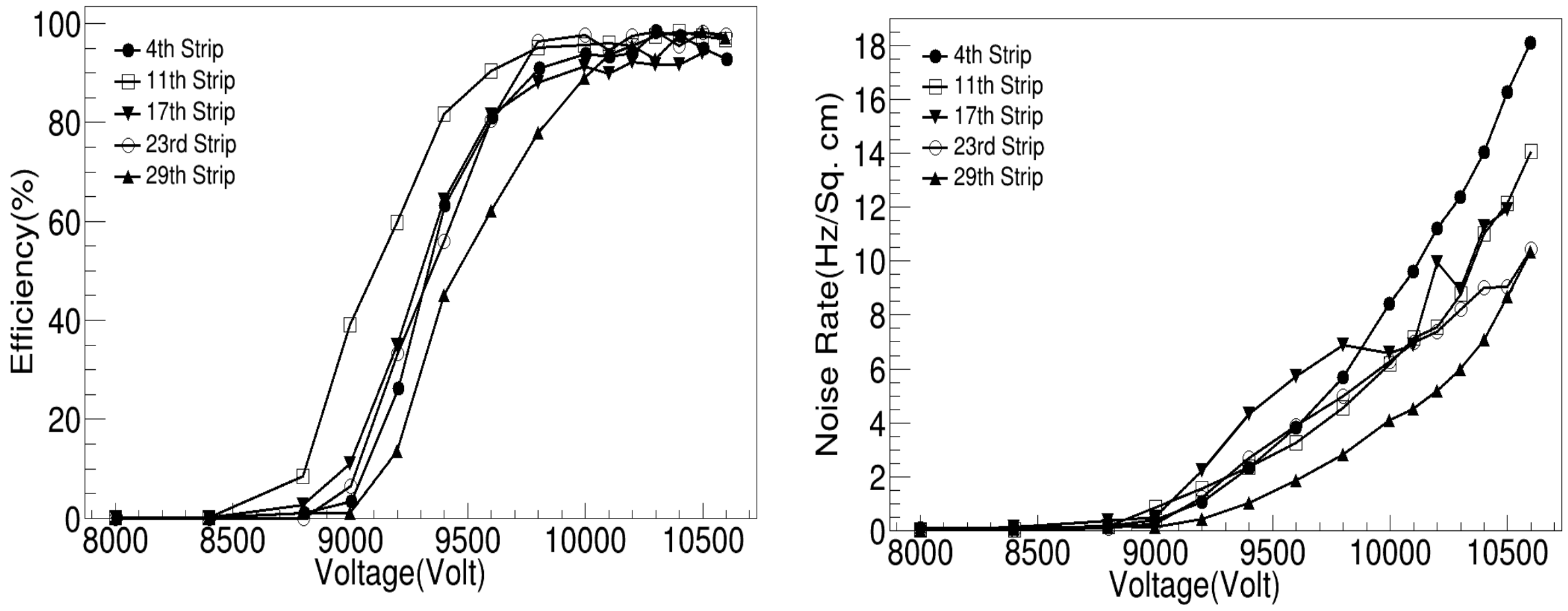

Figure 9: Efficiency vs. voltage characteristics (left), noise rate vs. voltage characteristics (right) for 5 different RPC strips for the 1st gas composition.

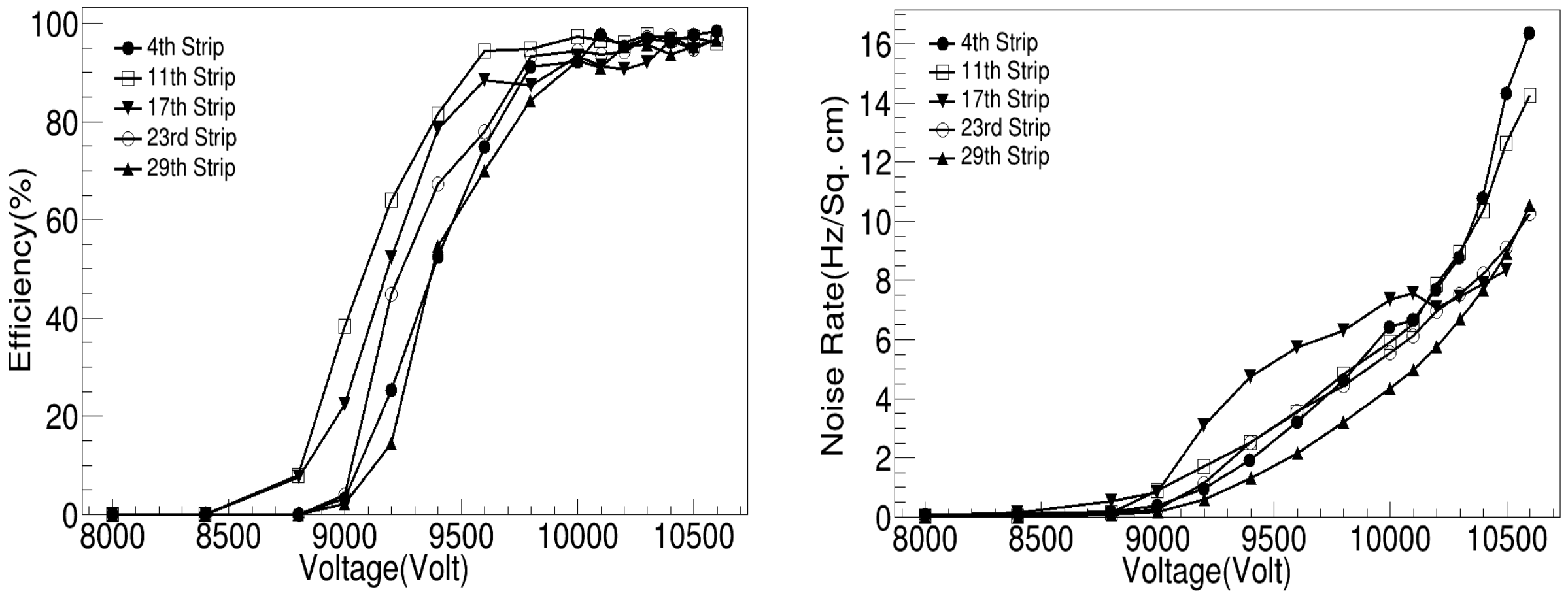

Figure 10: Efficiency vs. voltage characteristics (left), noise rate vs. voltage characteristics (right) for 5 different RPC strips for the 2nd gas composition.

## 5. Conclusions

The presented work establishes the operating parameters for a glass based RPC for different gas compositions. We have performed the leakage current studies for varying gas compositions, the efficiency and noise rate studies for the fixed gas flow rate keeping the temperature and relative humidity controlled throughout the measurements. We have also performed the efficiency and noise rate studies at 5 different locations on the RPC with the same set of parameters. From the leakage current measurements, it is concluded that RPC can be operated safely for the 1st and the 2nd gas compositions. On the basis of the efficiency and noise rate studies, it is concluded that the glass RPC shows a stable performance for the 1st and the 2nd gas compositions. The efficiency and noise rate characteristics have been found to be uniform throughout the RPC for the two selected gas compositions.

## Acknowledgements

We would like to thank Department of Science and Technology (DST) and Department of Atomic Energy (DAE)/Govt. of India for providing the financial support to accomplish this work. We would also like to acknowledge the help of other INO collaborators in procuring some of the raw materials for the detector assembly.